\documentclass[aps,prl,twocolumn,groupedaddress,showpacs]{revtex4-1}

\usepackage{graphicx}
\usepackage{color}

\begin{document}


\title{Solution to Big-Bang Nucleosynthesis in Hybrid Axion Dark Matter Model}


\author{Motohiko Kusakabe$^{1}$}
\email{kusakabe@icrr.u-tokyo.ac.jp}
\author{A.B. Balantekin$^{2,3}$}
\email{baha@physics.wisc.edu}
\author{Toshitaka Kajino$^{3,4}$}
\email{kajino@nao.ac.jp}
\author{Y. Pehlivan$^{3,5}$}
\email{ypehlivan@me.com}
\affiliation{
$^1$Institute for Cosmic Ray Research, University of Tokyo, Kashiwa, Chiba 277-8582, Japan}
\affiliation{
$^2$Department of Physics, University of Wisconsin, Madison, WI 53706, USA}
\affiliation{
$^3$National Astronomical Observatory of Japan 2-21-1 
Osawa, Mitaka, Tokyo, 181-8588, Japan}
\affiliation{
$^4$Department of Astronomy, Graduate School of Science, 
University of Tokyo, 7-3-1 Hongo, Bunkyo-ku, Tokyo, 113-0033, Japan}
\affiliation{
$^5$Mimar Sinan Fine Arts University, Besiktas, Istanbul 34349, Turkey
}



\date{\today}

\begin{abstract}
Following a recent suggestion of axion cooling of photons between the nucleosynthesis and recombination epochs in the Early Universe, we investigate a hybrid model with both axions and relic supersymmetric particles. In this model we demonstrate that the $^7$Li abundance can be consistent with observations without destroying the important concordance of deuterium abundance. 
\end{abstract}


\maketitle


{\bf Introduction} -- Low-metallicity halo stars exhibit a plateau of  $^7$Li abundance, indicating the primordial origin 
of $^7$Li \cite{Spite:1982dd}. However, the amount of $^7$Li needed to be consistent with the cosmic microwave background observations \cite{Komatsu:2010fb} is significantly more than $^7$Li observed in old halo stars \cite{Asplund:2005yt}. (Even though $^7$Li can be both produced and destroyed in stars, old halo dwarf stars are expected to have gone through little nuclear processing). Recent improvements in the observational and experimental data seem to make the discrepancy worse \cite{Cyburt:2008kw,Coc:2011az}. One possible solution is to invoke either nuclear physics hitherto excluded from the Big-Bang Nucleosynthesis (BBN) \cite{Chakraborty:2010zj,Cyburt:2009cf} or new physics such as variations of fundamental couplings \cite{Coc:2006sx,Cheoun:2011yn}, and particles not included in the Standard Model \cite{Lindley1979MNRAS.188P..15L,Ellis:1984er,Dimopoulos:1987fz,Reno:1987qw,Terasawa:1988my,Kawasaki:1994af,Kawasaki:1994sc,Jedamzik:1999di,Kawasaki:2000qr,Cyburt:2002uv,Kawasaki:2004qu,Jedamzik:2004er,Jedamzik:2004ip,Ellis:2005ii,Jedamzik:2006xz,Kusakabe:2006hc,Kusakabe:2008kf,Cyburt:2010vz,Pospelov:2010cw,Pospelov:2006sc,Kohri:2006cn,Cyburt:2006uv,Jedamzik:2007cp,Pospelov:2007js,Jittoh:2007fr,Hamaguchi:2007mp,Kusakabe:2007fu,Kamimura:2008fx,Kusakabe:2010cb,Jittoh:2010wh,Bird:2007ge,Jedamzik:2009uy,Pospelov:2010hj}. Effects of massive neutral relic particles on BBN were extensively studied \cite{Lindley1979MNRAS.188P..15L,Ellis:1984er,Dimopoulos:1987fz,Reno:1987qw,Terasawa:1988my,Kawasaki:1994af,Kawasaki:1994sc,Jedamzik:1999di,Kawasaki:2000qr,Cyburt:2002uv,Kawasaki:2004qu,Jedamzik:2004er,Jedamzik:2004ip,Ellis:2005ii,Jedamzik:2006xz,Kusakabe:2006hc,Kusakabe:2008kf,Cyburt:2010vz,Pospelov:2010cw}. 

More recently an alternative solution to the lithium puzzle was proposed. Since the last photon scattering occurs after the end of the nucleosynthesis, one can search for a mechanism for the cooling of photons before they decouple. It was suggested that dark matter axions could form a Bose-Einstein condensate (BEC) \cite{Sikivie:2009qn,Erken:2011dz}. Such a condensate would cool the photons between the end of BBN and epoch of photon decoupling, reducing the baryon-to-photon ratio WMAP infers,  as compared to its BBN value \cite{Erken:2011vv}. An alternative mechanism for such a cooling is resonant oscillations between photons and light abelian gauge bosons in the hidden sector 
\cite{Jaeckel:2008fi}. There are two {\it prima facie} problems with the axion BEC-photon cooling hypothesis: it overpredicts primordial deuterium (D) abundance as well as the effective number of neutrinos. Even though D is easy to destroy, one does not expect the sum of abundances of D and $^3$He to change significantly in the course of cosmic evolution \cite{Steigman:1995kuIocco:2008va}. Hence it is important to find a parameter region in which predicted abundances of D and $^7$Li are consistent with observations. In this letter we demonstrate the existence of such a parameter region using a model with axions and massive relic particles.

\begin{figure}
\begin{center}
\includegraphics[width=7.5cm,clip]{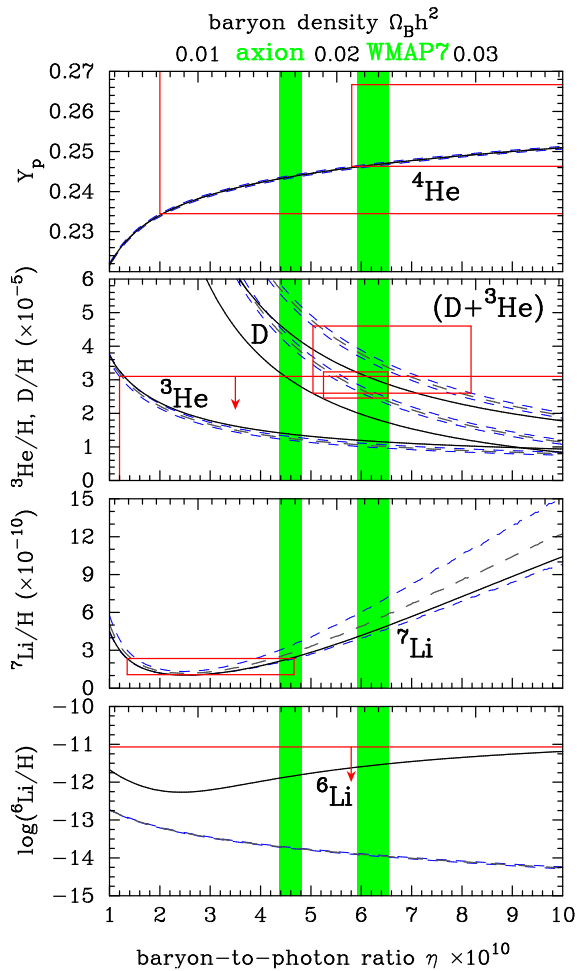}
\caption{ Abundances of $^4$He (mass fraction), D, $^3$He,
 $^7$Li and $^6$Li (number ratio relative to H) as a function of the
 baryon-to-photon ratio $\eta$ or the baryon energy density parameter $\Omega_B
 h^2$ of the universe.  The thick dashed curves are for SBBN.  
The thin dashed curves around them show the regions of $95 \%$ C. L. in accordance with the nuclear reaction rate uncertainties. The boxes correspond 
to the adopted abundance constraints on the SBBN model.  
The vertical stripes represent the 2$\sigma$ limits on $\Omega_B h^2$ or $\eta$ for the SBBN model (taken from the constraint by WMAP~\cite{Larson:2010gs} and labeled as WMAP7) and for the axion BEC model (labeled as axion). The solid curves are the results obtained with the long-lived decaying particle model with parameters fixed to ($\tau_X$, $\zeta_X$)=($10^6$ s, $2\times 10^{-10}$ GeV) (see text).\label{fig1}}
\end{center}
\end{figure}


{\bf The Hybrid Model} --  We carried out BBN network calculations using Kawano's code~\cite{Kawano1992,Smith:1992yy} by including Sarkar's correction for $^4$He abundances~\cite{Sarkar:1995dd}. JINA REACLIB Database V1.0 \cite{Cyburt2010} is used for light nuclear ($A \le  10$) reaction rates including uncertainties together with data \cite{Descouvemont2004,Ando:2005cz,Cyburt:2008up}. Adopted neutron lifetime is $878.5 \pm 0.7_{\rm stat} \pm 0.3_{\rm sys}$~s~\cite{Serebrov:2010sg} based on improved measurements \cite{Serebrov:2004zf}. Taking into account the uncertainties in the rates of twelve important reactions in BBN~\cite{Smith:1992yy}, we employ regions of $95\%$ C. L. in our calculations.

We compare our results with the abundance constraints from observations. 
For the primordial D abundance, the mean value estimated from Lyman-$\alpha$ absorption systems in the foreground of high redshift quasi-stellar objects is $\log({\rm D}/{\rm H})=-4.55\pm 0.03$~\cite{Pettini2008}. We adopt this value together with a $2\sigma$ uncertainty, i.e., $2.45\times10^{-5}< {\rm D}/{\rm H}< 3.24\times10^{-5}$.
$^3$He abundance measurements in Galactic HII regions through the $8.665$~GHz
hyperfine transition of $^3$He$^+$ yield a value of $^3$He/H=$(1.9\pm 0.6)\times
10^{-5}$~\cite{Bania:2002yj}. Although 
the constraint should be rather weak considering
its uncertainty, we take a $2\sigma$ upper limit and adopt $^3{\rm He}/{\rm H}< 3.1\times 10^{-5}$.

We also utilize a limit on the sum of primordial abundances of D and $^3$He taken from an abundance for the protosolar cloud determined from observations of solar wind, i.e., (D+$^3$He)/H=$(3.6\pm0.5)\times 10^{-5}$ \cite{1998SSRv...84..239G}.  This abundance can be regarded as constant at least within the standard cosmology since it is not affected by stellar activities significantly despite an effect of D burning into $^3$He via $^2$H($p,\gamma$)$^3$He would exist~\cite{Steigman:1995kuIocco:2008va}.

For the primordial $^4$He abundance, we adopt two different constraints from recent reports: $Y_{\rm p}=0.2565\pm 0.0051$~\cite{Izotov:2010ca} and $Y_{\rm p}=0.2561\pm 0.0108$~\cite{Aver:2010wq} both of which are derived from observations of metal-poor extragalactic
HII regions. Adding $2\sigma$ uncertainties leads to $0.2463 < Y_{\rm p} < 0.2667$~\cite{Izotov:2010ca} and $0.2345 < Y_{\rm p} < 0.2777$~\cite{Aver:2010wq}. 

$^6$Li plateau of metal-poor halo stars (MPHSs), yields the upper limit of 
$^6$Li/H$=(7.1\pm 0.7)\times 10^{-12}$~\cite{Asplund:2005yt}. Adding a $2\sigma$ uncertainty, we adopt $^6{\rm Li/H} < 8.5\times 10^{-12}$. 

For the $^7$Li abundance, we adopt the limits $\log(^7$Li/H)$=-12+(2.199\pm 0.086)$ (with $95\%$ C. L.) derived from recent observations of MPHSs in the 3D nonlocal thermal equilibrium model~\cite{Sbordone2010}, i.e. $1.06\times 10^{-10} < {\rm ^7Li/H} < 2.35\times 10^{-10}$. 

Figure \ref{fig1} shows the abundances of $^4$He ($Y_{\rm p}$; mass fraction), D, $^3$He, $^7$Li and $^6$Li (number ratio relative to H) as a function of the baryon-to-photon ratio $\eta$ or the baryon energy density parameter $\Omega_B h^2$ of the universe, where $h$ is the Hubble constant in units of 100 km/s/Mpc.  The thick dashed curves are the results of the standard BBN (SBBN) with a neutron lifetime of $878.5\pm 0.8$~s. Thin dashed curves around them show regions of $95 \%$ C. L. from uncertainties in the nuclear reaction rates. The boxes represent adopted abundance constraints as summarized above. The vertical stripes correspond to the 2$\sigma$ limits on $\Omega_B h^2$ or $\eta$. The values provided by WMAP~\cite{Larson:2010gs} (labeled WMAP7) are 
\begin{equation}
\label{WMAP7 values}
\Omega_B h^2=0.02258^{+0.00114}_{-0.00112} \quad \eta=(6.225^{+0.314}_{-0.309})\times 10^{-10}.
\end{equation}
Values predicted by the BEC model (labeled axion) are smaller by a factor of $(2/3)^{3/4}$ at the BBN epoch~\cite{Erken:2011vv}: 
\begin{equation}
\label{BEC values}
\Omega_B h^2=0.01666^{+0.00084}_{-0.00083} \quad \eta=(4.593^{+0.232}_{-0.228})\times 10^{-10}.
\end{equation}

It can be seen that the adoption of the $\eta$ value from WMAP leads to a $^7$Li abundance calculated in the axion BEC model, which is in reasonable agreement with the observations.  However, we lose the important consistency in D abundance.  Ref. \cite{Erken:2011vv} noted that astronomical measurements of primordial D abundance can have a significant uncertainty as well as a possibility that D is burned by nonstandard stellar processes.  Even if their assumption were true, stellar processes are not expected to change the sum of D and $^3$He abundances~\cite{Steigman:1995kuIocco:2008va}.  As seen in Fig. \ref{fig1}, the constraint on (D+$^3$He)/H abundance seems to exclude the original axion BEC model.  Ultimately, this model is viable only when the abundance of (D+$^3$He)/H is reduced through some exotic processes.

It is known that nonthermal photons can be generated through electromagnetic energy injections by the radiative decay of long-lived particles after the BBN epoch~\cite{Ellis:1984er,Kawasaki:1994sc}.  These nonthermal photons can photodisintegrate background light elements~\cite{Lindley1979MNRAS.188P..15L,Ellis:1984er,Cyburt:2002uv,Kawasaki:2004qu,Jedamzik:2006xz,Kusakabe:2006hc}.  We adopt the method of  Ref.~\cite{Kusakabe:2006hc} to calculate the nonthermal nucleosynthesis, where we incorporated new thermal reaction rates as described above.  In addition, we adopt updated reaction rates of $^4$He photodisintegration~\cite{Kusakabe:2008kf} derived from the cross section data using precise measurements with laser-Compton photons~\cite{Shima:2005ix,Kii:2005rk}.  Effects of electromagnetic energy injection depend on two parameters.  One is $\zeta_X=(n_X^0/n_\gamma^0)E_{\gamma 0}$ where $(n_X^0/n_\gamma^0)$ is the number 
 ratio of the decaying particle $X$ and the background radiation before the decay of $X$, and $E_{\gamma 0}$ is the energy of photon emitted at the radiative decay.  The other is $\tau_X$, the lifetime of the $X$ particle.

\begin{figure}
\begin{center}
\includegraphics[width=8.6cm,clip]{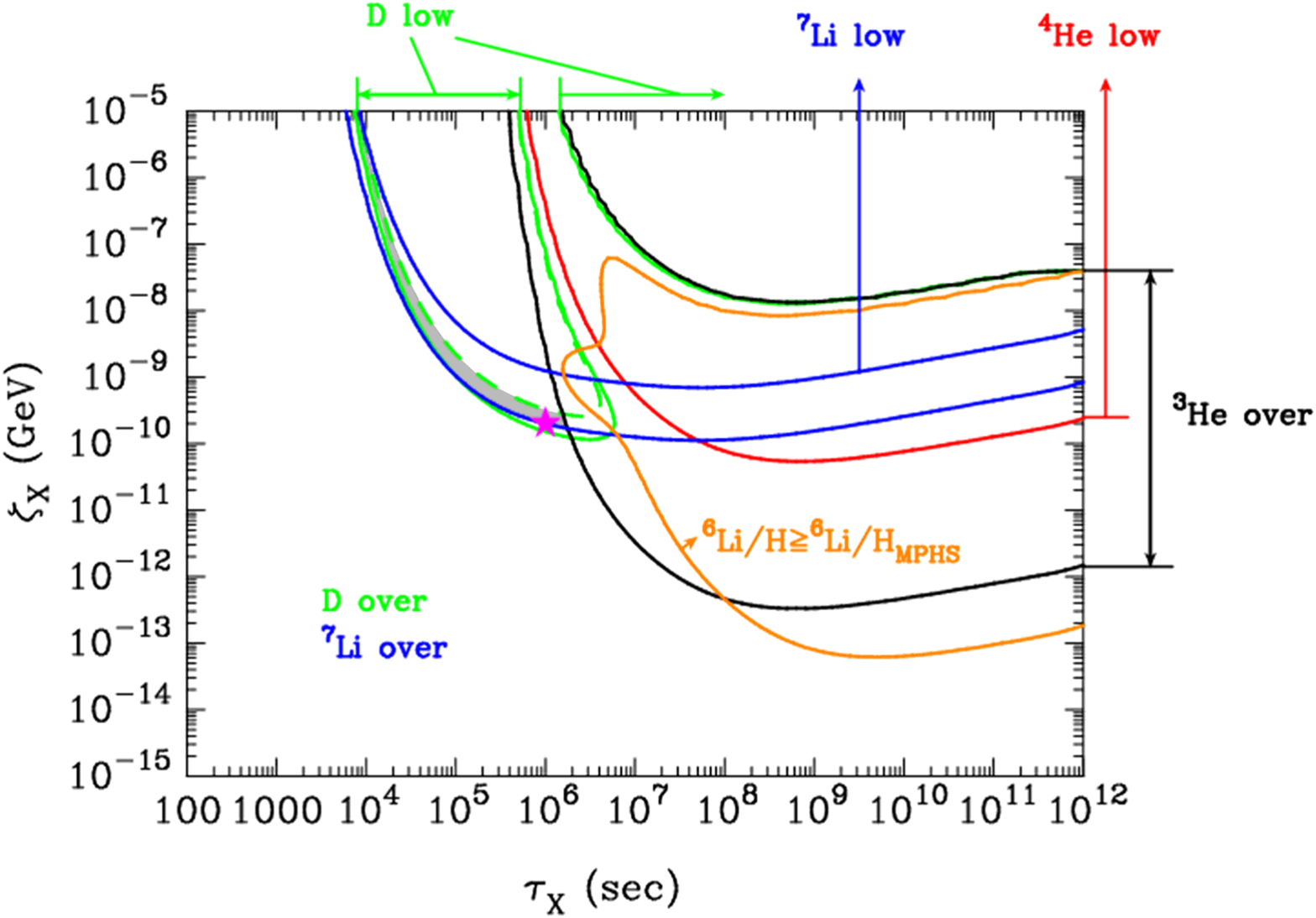}
\caption{ Parameter space of the hybrid model
 $(\tau_X,\zeta_X)$ for the value of $\eta=4.6\times 10^{-10}$ provided by the axion BEC model.
 The contours identify the regions where the nuclei are overproduced or underproduced
 (``over'' and ``low'', respectively) with respect to the adopted abundance constraints.
 $^4$He mass fraction (red line) and
 $^3$He/H (black lines),
 D/H (green solid and dashed lines for upper and lower limits, respectively), and $^7$Li/H (blue line) number ratios are
 shown. The orange line is the contour of $^6$Li/H=$7.1\times 10^{-12}$. 
 In the gray-colored region all abundances are within the limits of observational constraints.} \label{fig2}
\end{center}
\end{figure}

Figure \ref{fig2} shows the parameter space in our hybrid model.  For $^4$He, we adopt the conservative constraint with larger uncertainty~\cite{Aver:2010wq}.  We also show the contour for $^6$Li production at the observed level, i.e., $^6$Li/H$=7.1 \times 10^{-12}$.  This figure shows the result of nonthermal nucleosynthesis induced by the radiative decay of long-lived particles with the $\eta$ value of the axion BEC model [Eq. (\ref{BEC values})].  Except for D and $^7$Li, contours are similar to those represented in Ref~\cite{Kusakabe:2006hc} where BBN epoch $\eta$ value is assumed to be the same as WMAP $\eta$ value [Eq.~(\ref{WMAP7 values})].


In the very small colored region, calculated primordial abundances of all nuclides including D and $^7$Li are simultaneously in ranges of adopted observational constraints. We conclude that the present model eliminates the main drawback of the original axion BEC model by reducing primordial D abundance via $^2$H($\gamma$,$n$)$^1$H reaction, where $\gamma$'s are nonthermal photons.  We note that the decaying particle model with the WMAP $\eta$ value cannot resolve the $^7$Li problem by itself~\cite{Ellis:2005ii,Kusakabe:2006hc}.

The effect of the radiative decay on other elemental abundances is not significant except for $^7$Li.  Since energetic photons produced quickly collide with background photons and create $e^+e^-$ pairs, nonthermal photon spectra developed by the decay has a cutoff energy $E_{\rm C}=m_e^2/22T$ where $m_e$ is the electron mass~\cite{Kawasaki:1994sc}.  The decay at earlier (hotter) universe then triggers nonthermal photons with lower cutoff energies.  The threshold energies of $^7$Be and D photodisintegration, $^7$Be+$\gamma\rightarrow ^3$He+$^4$He and D+$\gamma \rightarrow n +p$ are 1.5866 MeV and 2.2246 MeV, respectively.  These two nuclei are very fragile against photodisintegration.  When the decay occurs early at relatively high $T_9\equiv T/(10^9~{\rm K}) \gtrsim 10^{-2}$ which corresponds to $\tau_X\lesssim 10^6$~s, nonthermal photon spectra contain photons to dissociate $^7$Be and D, while keeping other nuclides intact.  The gray region indicates parameters which result in moderate destruction of D which is overproduced in the original BEC model because of low $\eta$.  Above that region, D is destroyed too much by photodisintegration, while below it D abundance is too high. 

\begin{figure}
\begin{center}
\includegraphics[width=8.0cm,clip]{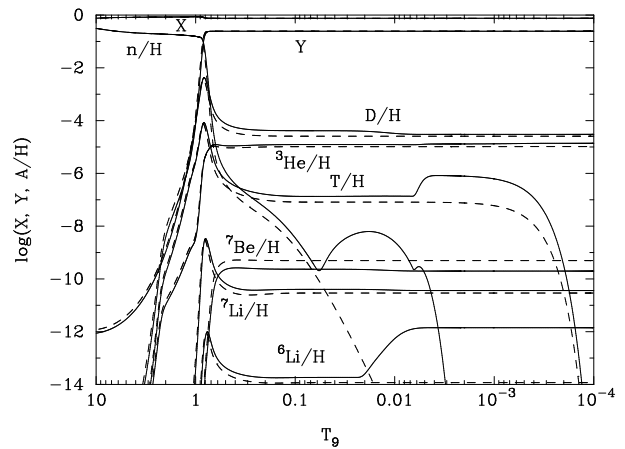}
\caption{ Mass fractions of H and $^4$He ($X_{\rm p}$ and $Y_{\rm p}$, respectively) and number ratios of other nuclides relative to H as a function of $T_9$. Solid lines show the abundances calculated in the hybrid model with the parameters ($\tau_X$, $\zeta_X$)=($10^6$ s, $2\times 10^{-10}$ GeV) which correspond to the point  indicated with a star in Fig. \ref{fig2}. The dashed lines show the SBBN prediction.} \label{fig3}
\end{center}
\end{figure}

We next present the results of a BBN calculation in our hybrid model with a fixed set of parameters given by ($\tau_X$, $\zeta_X$)=($10^6$ s, $2\times 10^{-10}$ GeV) and the $\eta$ value given in Eq. (\ref{BEC values}). This corresponds to the point indicated with a star in Fig. (\ref{fig2}) and yields the required D and $^7$Li abundances as seen in Figure \ref{fig3}. This figure shows H and $^4$He mass fractions (denoted by $X_{\rm p}$ and $Y_{\rm p}$), and $n$, D, T, $^3$He, $^6$Li, $^7$Li and $^7$Be number ratios relative to H as a function of the temperature. The abundances calculated using the axion BEC + long-lived decaying particle model with the parameters ($\tau_X$, $\zeta_X$)=($10^6$ s, $2\times 10^{-10}$ GeV) and the $\eta$ value provided by BEC model [Eq. (\ref{BEC values})] are shown in solid lines whereas the SBBN prediction is plotted by dashed lines.  The small difference at $T_9\gtrsim 0.06$ observed between solid and dashed lines is caused by difference between initial $\eta$ values.  At $0.06\gtrsim T_9 \gtrsim 7\times 10^{-3}$ (corresponding to the cosmic time of $t\sim 5\times 10^4$--$4\times 10^6$~s), effects of $^2$H($\gamma,n$)$^1$H are seen in the decrease of D and the increase of $n$ abundances.  We find a slight decrease in $^7
 $Be abundance.  This is caused through reactions $^7$Be($\gamma,^3$He)$^4$He (threshold energy: 1.5866 MeV), $^7$Be($\gamma,p$)$^6$Li (5.6858 MeV), and $^7$Be($\gamma,2pn$)$^4$He (9.3047 MeV).  The second reaction increases the $^6$Li abundance.  Finally, at $T_9\lesssim 7\times 10^{-3}$, where the abundance of long-lived $X$ particle is already less than 3 \% of the initial abundance, effect of $^4$He photodisintegration is to increase $^3$H and $n$ abundances.


In Fig. \ref{fig1}, solid lines show the results for the SBBN + long-lived decaying particle model with the same parameter values, i.e., ($\tau_X$, $\zeta_X$)=($10^6$ s, $2\times 10^{-10}$ GeV).  Obviously the abundances of D and $^7$Li (produced partly as $^7$Be) are reduced, while that of $^6$Li is increased from those of SBBN.

{\bf Conclusions} --  We used a hybrid axion and massive relic particle model in which axions cool the photons and relic particles produce non-thermal photons to eliminate the high D abundance in the original axion BEC model. Our hybrid model also produces $^6$Li keeping $^7$Li abundance at the level of Population II Spite plateau. Our work thus demonstrates that the $^7$Li abundance can be consistent with observations without destroying the important concordance of deuterium abundance. 

\begin{acknowledgments}
This work was supported in part by Grants-in-Aid for Scientific Research of the JSPS (200244035) and for Scientific Research on Innovative Area of MEXT (20105004), in part by JSPS Grant No.21.6817, in part by the U.S. National Science Foundation Grant No. PHY-0855082, in part by the Council of Higher Education of Turkey, and 
in part by the University of Wisconsin Research Committee with funds
granted by the Wisconsin Alumni Research Foundation.
\end{acknowledgments}


%

\end{document}